# Gate Tunable Magneto-resistance of Ultra-Thin WTe$_2$ Devices

Xin Liu[1*], Zhiran Zhang[1], Chaoyi Cai[1], Shibing Tian[1], Satya Kushwaha[2], Hong Lu[1], Takashi Taniguchi[3], Kenji Watanabe[3], Robert J. Cava[2], Shuang Jia[1,4*], Jian-Hao Chen[1,4*]

[1] International Center for Quantum Materials, Peking University, Beijing 100871, China

[2] Department of Chemistry, Princeton University, Princeton, USA

[3] High Pressure Group, National Institute for Materials Science, 1-1 Namiki, Tsukuba, Ibaraki 305-0044, Japan

[4] Collaborative Innovation Center of Quantum Matter, Beijing 100871, China

[*]Xin Liu (liux23@pku.edu.cn), Shuang Jia (jiashuang@gmail.com) and Jian-Hao Chen (chenjianhao@pku.edu.cn)

*Abstract: In this work, the magneto-resistance (MR) of ultra-thin WTe$_2$/BN heterostructures far away from electron-hole equilibrium is measured. The change of MR of such devices is found to be determined largely by a single tunable parameter, i.e. the amount of imbalance between electrons and holes. We also found that the magnetoresistive behavior of ultra-thin WTe$_2$ devices is well-captured by a two-fluid model. According to the model, the change of MR could be as large as 400,000%, the largest potential change of MR among all materials known, if the ultra-thin samples are tuned to neutrality when preserving the mobility of 167,000 $cm^2V^{-1}s^{-1}$ observed in bulk samples. Our findings show the prospects of ultra-thin WTe$_2$ as a variable magnetoresistance material in future applications such as magnetic field sensors, information storage and extraction devices, and galvanic isolators. The results also provide important insight into the electronic structure and the origin of the large MR in ultra-thin WTe$_2$ samples.*

**Introduction**

1T'-Tungsten ditelluride (WTe$_2$) is a layered transition metal dichalcogenide (TMDC) with a distorted structure that preserves inversion symmetry in the out-of-plan direction, contrasting with TMDCs with 2H phases, such as 2H-MoS$_2$ [1]. The material in its pristine bulk form is a semimetal [2, 3]. It exhibits rich physics such as extraordinarily large and non-saturating magneto-resistance (XMR) [4], superconductivity under high pressure [5, 6], and may be a type-II Weyl semimetal (WSM) at a particular level of electron doping [7-10]. Furthermore, reports on thin films of WTe$_2$ show the tunability of magnetoresistance revealing interesting new phenomena such as the transition from weak anti-localization to weak localization [11], the depletion of holelike carriers in the suppressed-MR regime [12], long-range field effect [13], the topological insulator-like behavior [14], and negative longitudinal MR indicating WTe$_2$ a type-II WSM [15]. The origin of the XMR in WTe$_2$, together with its potential application in magnetic field sensing and in information storage, has attracted much attention in the scientific and technical community [16-21]. The majority of the research in XMR has been carried-out in bulk $WTe_2$ samples with near-perfect electron-hole compensation, and the results support the picture that the XMR arises from such near-perfect compensation of electrons and holes [17], i.e. a two-fluid picture. However, electrolyte gating experiments on thinner $WTe_2$ samples (~70nm) have shown non-saturating XMR that deviates from the two-fluid theory [18]. Here, we have carried out careful experiments with solid-dielectric gated ultra-thin $WTe_2$ samples (~10nm) that are far away from charge neutrality. We find that in this regime, the MR of the samples can still be well explained by the two-fluid model, and the sample shows 2D weak anti-localization effects at low temperatures. We also found that the change of MR of the ultra-thin $WTe_2$ is determined largely by the density difference between the electron and hole carriers, pointing to possible future application of this material in electric-field tunable, variable sensitivity magnetic field sensors [22-24].

**Device fabrication and measurement**

The ultra-thin $WTe_2$ samples measured in this letter are mechanically exfoliated from bulk $WTe_2$ crystal and transferred on to thin h-BN single crystals placed on 300nm $SiO_2$/Si substrates [25]. We found that using single crystal BN substrates resulted in an increase in the mobility of our samples which are 10nm thick or less (see section S4 at Supporting information for details). $WTe_2$ bulk crystals are synthesized using chemical vapor transport technique [4] and h-BN bulk crystals are grown by the method described in ref. [26]. The thin h-BN (thickness～15nm) surface is free of dangling bonds, greatly alleviates the influence of surface charge traps in the $SiO_2$ layer, and could substantially improve quality of low-dimensional devices. Standard electron-beam lithography technique is used to pattern electrodes, consisting of 6nm Cr and 60nm Au, on the $WTe_2$ samples to form multi-terminal field effect devices (FEDs). We have taken particularly careful measures to ensure that the samples do not expose to ambient conditions at all. The sample preparation process, device fabrication process and electrical transport measurement are done in inert atmosphere, or with the sample capped with a protection layer. The protection layer consists of 200 nm thick polymethyl methacrylate (PMMA) or a bilayer of 200 nm PMMA and 200 nm MMA. Electrical- and magneto-transport measurements were carried out in a Quantum Design PPMS-9 with standard lock-in technique.

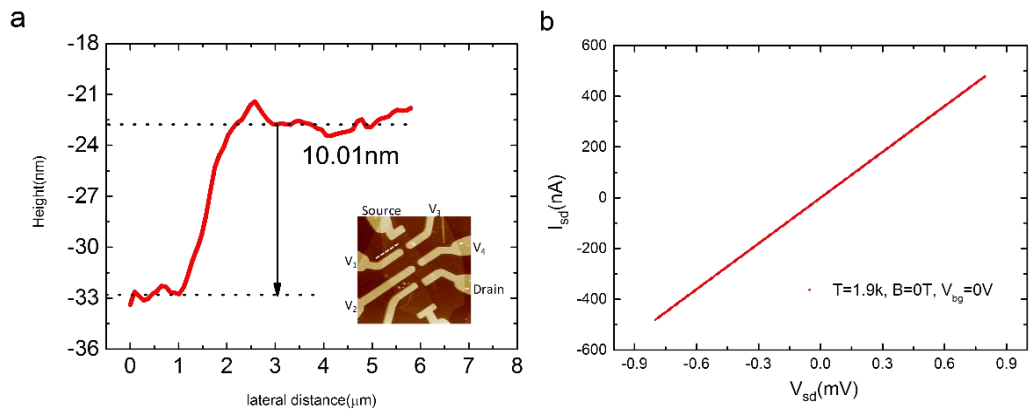

**Figure 1** Few-layer WTe$_2$ FET device characteristics. **(a)** Thickness of the WTe$_2$ FET device measured by atomic force microscopy (AFM). Inset shows the AFM micrograph of the device. The channel length (between $V_1$ and $V_2$) and width are 3.9 µm and 1 µm, respectively. The white dashed line in the inset panel marks the cross-section of the sample in (a). **(b)** Two-terminal $I_{sd}$-$V_{sd}$ characteristics of the device at T=1.9K without magnetic field and at zero back gate voltage.

We shall focus our discussion on one device (sample A) in the main text; data for other WTe$_2$ devices are shown in supporting information. An atomic force microscopy (AFM) micrograph of a 10nm-thick WTe$_2$ device as well as its height section profile is shown in figure 1a. Figure 1b shows the *I-V* characteristics of the WTe$_2$ field effect device at 1.9 Kelvin under zero magnetic field and zero back gate voltage, measured with two-probe configurations (between source and drain electrodes shown in the inset of figure 1a). The source-drain current ($I_{sd}$) varies quite linearly with the applied voltage $V_{sd}$ from -0.8mV to 0.8mV, with resistivity of $4.26 \times 10^{-4}$ Ω·cm, indicating an Ohmic contact to a metallic sample. Raman spectra of the device (see section S3 in supporting information) obtained in a Horiba Jobin Yvon LabRam HR Evolution system after the transport experiments showed that the WTe$_2$ sample is indeed in the 1T' phase [27] and has not degraded.

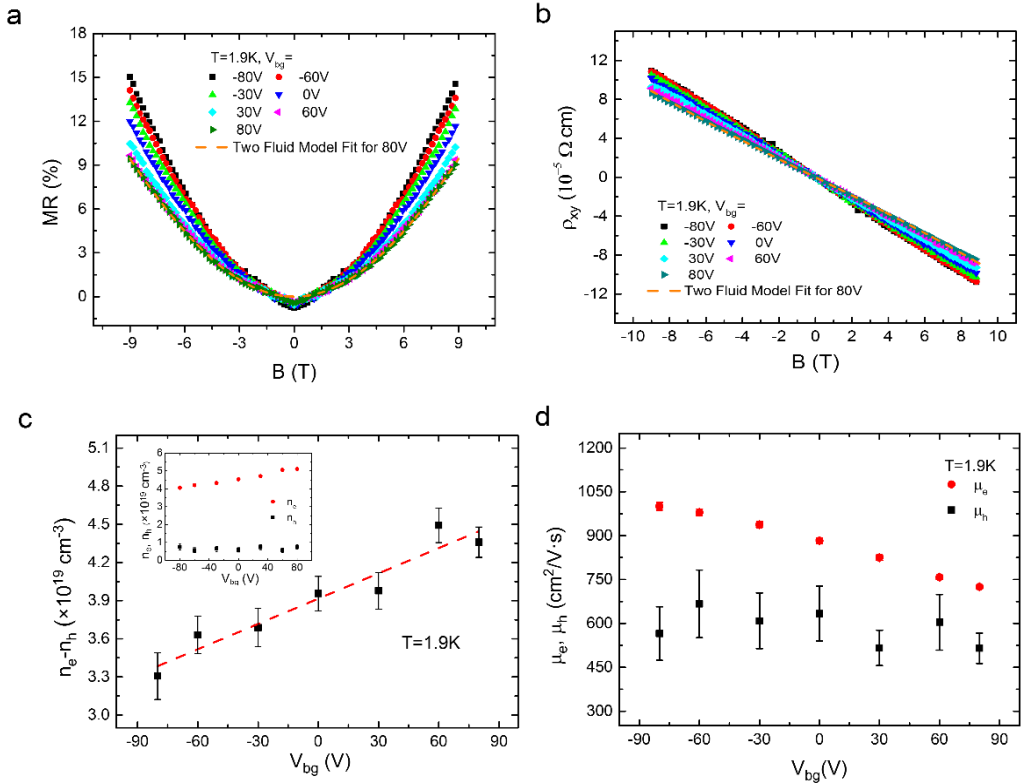

**Figure 2** Gate-dependent behavior in WTe$_2$ thin flake. **(a), (b)** Gate dependence of MR and $\rho_{xy}$ at T=1.9K. The colors mark different back gate voltage. Dashed lines are fits to the two-fluid model for MR and $\rho_{xy}$ at $V_{bg}$=80V. **(c), (d)** Gate dependence of net carrier density ($n_e$-$n_h$) and mobility for electrons and holes. Dashed line in (c) is a linear fit to the data. Inset

figure in (c) shows electron and hole density as a function of back gate voltage, respectively.

**Magneto-transport mechanism of WTe$_2$ FET devices**

Figures 2a and 2b show the longitudinal magneto-resistance $MR(B) = (\rho_{xx}(B) - \rho_{xx}(B=0))/\rho_{xx}(B=0)$ and Hall resistance $\rho_{xy}$ of the sample at $T$=1.9K, with the magnetic field applied perpendicular to the sample surface (along the c axis), and at different back gate voltages. All measurements of magnetoresistance are done in four-probe configurations to eliminate contact effects. The sample exhibits a positive $MR(B)$ and quasi-linear $\rho_{xy}(B)$ under all the gate voltages investigated in this experiment. For small magnetic field, $MR(B)$ has a cusp-shape which will be discussed later. For magnetic fields larger than 1 Tesla, $MR(B)$ can be fitted by a power law behavior, $MR \propto B^\gamma$, with the exponent $\gamma$ between 1.62 and 1.69 for different gate voltages, which is smaller than the exponent $\gamma$~1.94 observed in an 112nm-thick sample (see section S1 in supporting information), and smaller than the exponent $\gamma$~2 observed bulk crystals [4]. We will show later in this report that weak anti-localization should be taken into account (see figure 6) and the exponent $\gamma$ is not a constant of magnetic field for electron-hole uncompensated samples (see section S7 in supporting information).

It has been found by transport [28-30] and ARPES [16, 17, 31, 32] experiments that bulk WTe$_2$ crystals have 4-9 carrier pockets; however, researchers are just starting to examine how these pockets evolve as the sample is getting thinner [33]. Here we started out to analyze the data with the ansatz that there are two major types of carriers in the sample, one type is electrons and the other type is holes. We will show that this ansatz captures the majority of the physics in the high field magnetoresistance of ultra-thin WTe$_2$ samples, and that it is also consistent with low field magnetoresistance data.

In a two-fluid model [34], we have

$$\rho_{xx} = \frac{1}{e} \frac{(n_e u_e + n_h u_h) + (n_e u_h + n_h u_e) u_e u_h B^2}{(n_e u_e + n_h u_h)^2 + ((n_e - n_h) u_e u_h B)^2}$$

(1)

$$\rho_{xy} = \frac{1}{e} \frac{(n_e \mu_e^2 - n_h \mu_h^2) - (n_h - n_e) \mu_e^2 \mu_h^2 B^2}{(n_e u_e + n_h u_h)^2 + ((n_h - n_e) u_e u_h B)^2} B$$

(2)

$$\mathrm{MR} = \frac{\rho_{xx}(B) - \rho_{xx}(B=0T)}{\rho_{xx}(B=0T)} = \frac{\frac{n_h}{n_e}(u_e + u_h)^2 u_e u_h B^2}{\left(u_e + \frac{n_h}{n_e} u_h\right)^2 + \frac{n_h}{n_e}\left(\left(\frac{n_h}{n_e} - 1\right) u_e u_h B\right)^2}$$

(3)

where $n_e$ ($n_h$) and $u_e$ ($u_h$) are carrier density and mobility for electrons (holes), respectively. At all the gate voltages, both the $MR$ and $\rho_{xy}$ of the ultra-thin device can be simultaneously fitted by equation (3) and (2), and $n_e$, $n_h$, $u_e$, $u_h$ can be extracted from the fit. Using the least squares method, we determine the values of the four parameters with minimum error (see section S9 in supporting information for details).

Figure 2c shows the net charge carrier density $n=n_e-n_h$ as a function of $V_g$, and the dashed line is a linear fit to the data. The induced charge in the sample by the silicon back gate is: $ne = C_g \Delta V_g$, where $e$ is the elementary charge and $C_g$ is the parallel-plate capacitance of the device per unit area. Thus from the linear fit, we obtained a gate capacitance of $C_g = 1.062 \times 10^{-4} \, F/m^2$. Since the dielectric in our device consists of 15nm of h-BN (relative permittivity $\epsilon_{h-BN} \approx 3.5$) and 300nm of SiO$_2$ ($\epsilon_{SiO_2} \approx 3.9$), one can get the series capacitance for this multilayer system to be $C'_g = \left( \frac{1}{C_g^{h-BN}} + \frac{1}{C_g^{SiO_2}} \right)^{-1} = 1.089 \times 10^{-4} \, F/m^2$, in good agreement with our experimental data. The above analysis show that the longitudinal magnetoresistance of the device can be tuned electrostatically and that the phenomenological two-fluid model captures the main feature of the magneto-transport properties of our ultra-thin WTe$_2$ samples. Note that the ultra-thin samples in this study are predominately electron-doped, with electron densities 5-10 times larger than hole densities; in comparison, thicker samples (the 112nm-thick sample, see section S1 in supporting information) exfoliated from the same bulk crystal are found to be close to charge neutrality. The imbalance between electrons and holes in the ultra-thin devices are likely due to unintentional doping from the device fabrication process; such imbalance also allows us to access the highly electron-doped regime in ultra-thin WTe$_2$ samples to test the applicability of the two-fluid model [4, 18].

Figure 2d shows the gate-dependent mobility for electrons and holes. We note that the electron mobility decreases as the density becomes larger at T=1.9K, suggesting that charged impurities are not the dominating scattering source in this regime [35]. From figure 2d as well as from the inset of figure 2c, we find that $n_e$ and $u_e$ are being effectively tuned by the gate voltage while $n_h$ and $u_h$ are much less affected by $V_g$, which is likely caused by the fact that the density of states of electrons is much larger than that of holes in this highly electron-doped regime.

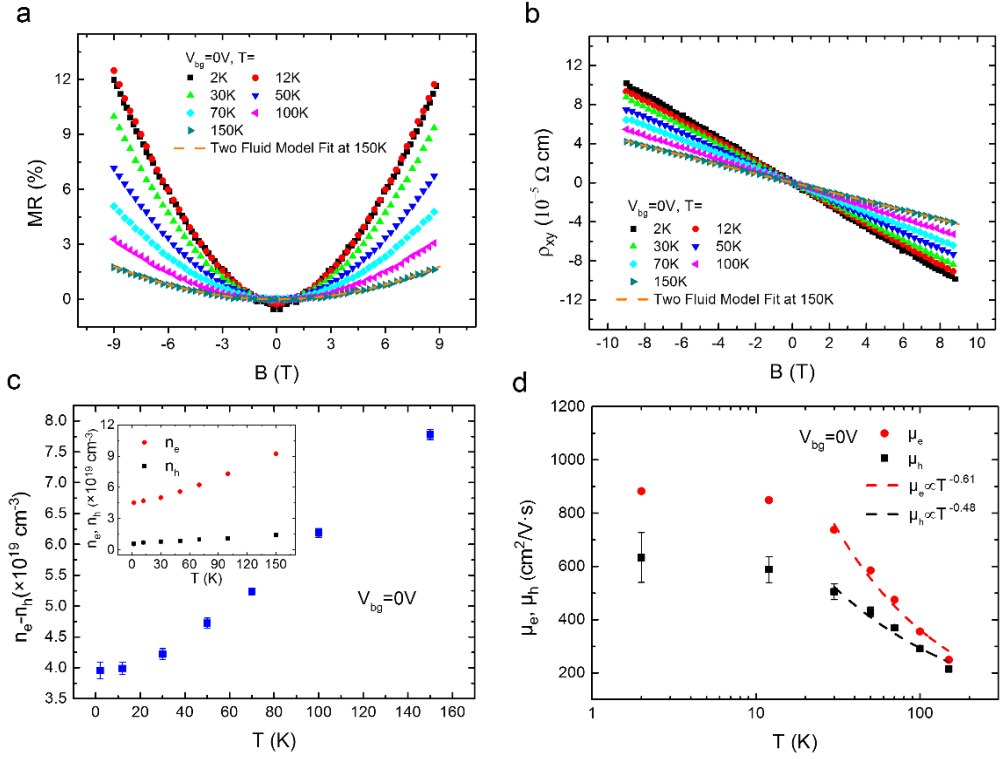

**Figure 3** Temperature-dependent behavior of a WTe$_2$ ultra-thin flake device. **(a)**, **(b)** *MR* and Hall resistivity as a function of magnetic field at different temperatures at $V_{bg}$=0V. Dashed lines are fit curve to the two-fluid model for *MR* and $\rho_{xy}$ at 150K. **(c)**, **(d)** Temperature dependence of net charge carrier density ($n_e$-$n_h$) and the mobility of electrons and holes at $V_{bg}$=0V. The inset panel in (c) shows electron and hole density vs. temperature. Dashed lines in (d) are power law fits for $\mu \propto T^{-\gamma}$, with the exponents $\gamma$≈0.61 and 0.48 for electrons and holes, respectively.

Figures 3a and 3b show the temperature-dependent *MR* and $\rho_{xy}$ of the device from *T*=1.9K to 150K and at zero gate voltage. The cusp-shape in *MR* at small magnetic field diminishes rapidly as the temperature increases, and will be discussed in detail later. At magnetic field larger than two Tesla, fitting the *MR* by a power law behavior $MR \propto B^\gamma$ results in a temperature-dependent exponent *γ* that changes from ~1.69 at *T*=1.9K to 2 at *T*=150K. At the meantime, $\rho_{xy}$ remains linear in B with its slope *k* changes monotonically at different temperatures. We are going to show in this letter that such temperature-dependence of *γ* and *k* is also well captured by the phenomenological two-fluid model.

Similar to the analysis of the gate dependent magnetoresistance, the *MR* and $\rho_{xy}$ of the device at different temperatures are fitted simultaneously by equation (3) and (2), and the dependencies of $n_e$, $n_h$, $u_e$, $u_h$ on temperature are extracted from the fit. Figure 3c shows the dependence of the net carrier charge $n_e$-$n_h$ as a function of temperature. It can be seen that by lowering the temperature, the sample rapidly tends to its charge neutrality on cooling from 150K to 50K; the trend slows down below 50K and saturates from 12K to 1.9K. The mobilities $u_e$ and $u_h$, on the other hand, increase following a power law of $\mu \propto T^{-\alpha}$ from 150K to 50K and then saturate from 12K to 1.9K. This suggests a connection between the decrease in $n_e$-$n_h$ and the increase of $u_e$ and $u_h$, which is consistent with the carrier density dependent measurements at a fixed temperature. Temperature-dependent movement of chemical potential has been seen in multiple semimetal bulk crystals [31, 36-38], and has been attributed to be the cause of a temperature-induced Lifshitz transition for $WTe_2$ bulk crystals [31]. Thus it is not surprising to see such temperature-dependent $n_e$-$n_h$ in ultra-thin $WTe_2$ samples. A fit to a power law behavior of the decreasing mobility with increasing temperature give an exponent $\alpha$ for electrons ($\alpha$=0.61) and holes ($\alpha$=0.48); these values are similar to those for few-layer black phosphorus [39] and dual-gated monolayer $MoS_2$ [40]. However, they are smaller than the theoretically predicted value ($\alpha \sim 1.52$) [41] and smaller than our experimental data obtained from bulk $WTe_2$ samples ($\alpha \sim 1.30$ to 1.51) (see section S2 in supporting information). The suppression of $\alpha$ is considered to be caused by a quenching of the characteristic homopolar mode in sandwiched ultra-thin device structures [40]. In the case of $WTe_2$, this means that ultra-thin samples can preserve their mobility, thus preserving their response to magnetic field, much better than their bulk counterparts at room temperature (see section S2 in supporting information), which is good for technological applications.

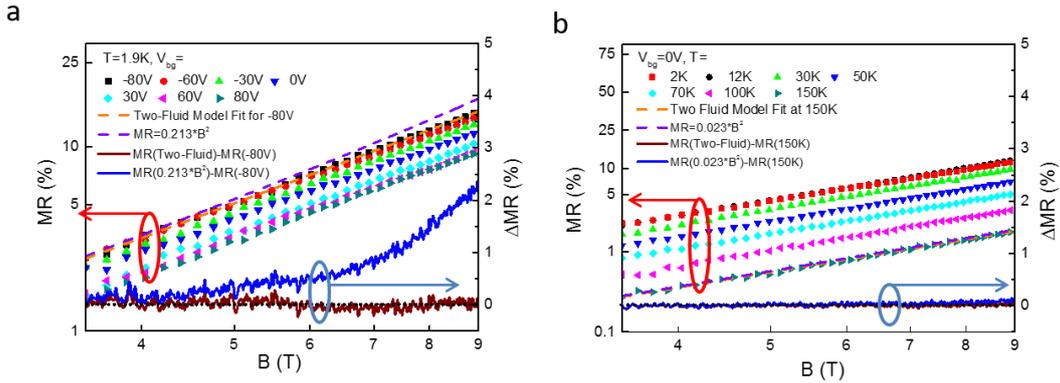

**Figure 4** **(a)** Gate-voltage dependent *MR* plotted on a log-log scale. The orange dashed lines are the two-fluid model fit to the *MR* at $V_g$ = -80 volts at *T*=1.9K, the purple dashed line is a $B^2$ *MR* curve as a guide to the eye; the blue solid line is the difference between the MR data at $V_g$ = -80 volts at *T*=1.9K and the $B^2$ behavior, while the wine solid line is the difference of the same data to the prediction of the two fluid model. **(b)** Temperature dependent MR plotted on a log-log scale. The orange dashed lines are the two-fluid model fit to the MR at $V_g$ = 0 volts at *T*=150K, the purple dashed line is a $B^2$ MR curve as a guide to the eye; the blue solid line is the difference between the MR data and the $B^2$ behavior, while the wine solid line is the difference between the same data and the prediction of the two fluid model.

Since the ultra-thin $WTe_2$ sample is not at the charge neutral point, we expect a saturation of magnetoresistance for high enough magnetic field, if the transport behavior of the sample follows the two-fluid model. Indeed, if we look closely into the *MR* curves, we confirm such saturation at high magnetic field. Figures 4a and 4b show the gate-voltage dependent *MR* and temperature dependent *MR* plotted in log-log scale. It can be seen that for all the gate voltages we applied at 1.9K, the *MR* of the sample deviates from the $\sim B^2$ functional form, while conforming to the prediction of the two-fluid model. In figure 4a, the blue solid line is the difference between the MR data at $V_g$ = -80 volts at $T$=1.9K and the $B^2$ behavior, while the wine solid line is the difference of the same data to the prediction of the two fluid model. A deviation from the $B^2$ behavior (e.g., the saturation of *MR* at high magnetic field) is clearly observed. At higher temperature, the mobility of the carriers drops much faster than the increase in the net charge density of the sample, leading to a higher saturation magnetic field, which is out of the range of our experimental apparatus. Thus the MR data at 150K fits the two-fluid model and a scaled $B^2$ functional form equally well (figure 4b).

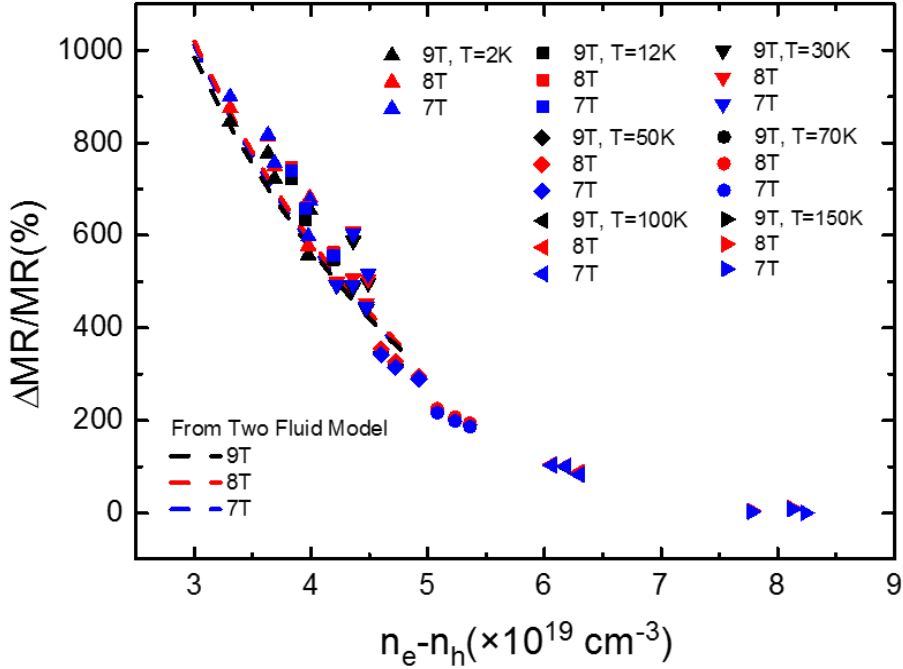

**Figure 5** Tunable MR by carrier density in a WTe$_2$ ultra-thin flake device. Charge carrier density dependence of normalized magneto-resistance, defined as *(MR(n)-MR(n$_0$))/MR(n$_0$)*, where $n = n_e-n_h$ is the net charge density and $n_0 = 8.2 \times 10^{19}$cm$^{-3}$ is the largest net charge density measured in this device. Dashed lines are the two-fluid model prediction of *MR(n$_e$-n$_h$)* curves for magnetic field of 7T, 8T and 9T.

Figure 5 is one of the primary observations of our study. It shows the change of *MR* as a function of $n_e$-$n_h$ at three different magnetic fields (7T, 8T, 9T) and at all temperatures from 150K to 1.9K. (The change of *MR* is defined as *(MR(n)-MR(n$_0$))/MR(n$_0$)*, where $n = n_e$-$n_h$ is the net charge density and $n_0 = 8.2 \times 10^{19}$cm$^{-3}$ is the largest net charge density measured in our experiment, at $T$ = 150K.) It can be seen that the change of *MR* increases monotonically as $n_e$-$n_h$ decreases, regardless of temperature and magnetic field. The two-fluid model predicted *MR(n$_e$-n$_h$)* curves for magnetic field of 7T, 8T and 9T are also shown in figure 4, showing similar insensitivity to the magnetic field applied. This property of ultra-thin $WTe_2$ devices is very useful in making future tunable sensitivity magnetic field sensors, where a universal dependence on a single parameter (net charge carrier) is preferred. It is worth noting that such a curve is empirical, and it is a useful derivation from the two-fluid model. Experimentally, we found that for the carrier mobility about 1000 $cm^2/Vs$ or lower, and for magnetic field 9T or lower, such single-parameter dependence of $\Delta MR$ on $n_e$-$n_h$ holds pretty well.

The largest change of *MR* measured in our experiment is 850%, in which the 2D electron-hole imbalance is tuned from $8.2 \times 10^{19} cm^{-3}$ to $3.2 \times 10^{19} cm^{-3}$. If we reached charge neutrality in this particular device, the change of MR could be 8,400% (see section S8 in supporting information for the calculation). Furthermore, as fabrication techniques improve, we expect the mobility of ultra-thin $WTe_2$ devices to finally approach that of bulk crystals. (Such rapid improvement of device fabrication techniques has been seen in the field of graphene, where it did not take a long time for the mobilities of graphene devices to improve from 10,000$cm^2/Vs$ [42] to 1,000,000$cm^2/Vs$ [25].) Using a fixed mobility value of $1.67 \times 10^5$ $cm^2V^{-1}s^{-1}$ from Ref. [43] for both *MR*(p/n=1) and *MR*(p/n=0.1559), we project a change of *MR* of 400,000%. Note that this should be a lower bound for the estimation, as the mobility at the neutrality point should be much higher than when p/n=0.1559 (see figure 2d and supporting information for more details). Thus we expect ultra-thin $WTe_2$ to be a very useful electric-field-tuned magnetoresistance material in future technological applications.

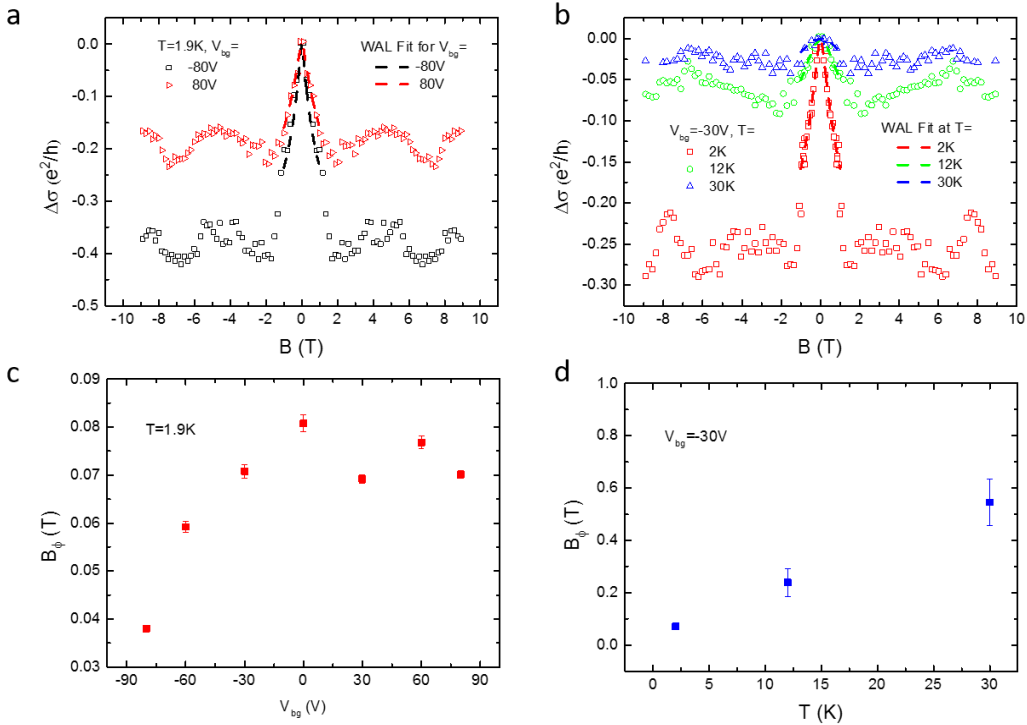

**Figure 6.** WAL in $WTe_2$ thin flake. **(a)** The change of conductance $\Delta\sigma_{xx}(B) = \sigma_{xx}(B) - \sigma_{xx}(B=0)$ of the ultra-thin $WTe_2$ device at $T$ = 1.9K at largest gate voltages $V_g$ = ±80 volts. **(b)** $\Delta\sigma_{xx}(B)$ curves for $T$ = 2K, 12K and 30K at gate voltage $V_g$ = -30 volts. The dashed lines in (a) and (b) are weak anti-localization fit to the data between ±1T of magnetic field. **(c)&(d)** Phase related field $B_\phi$ as a function of back gate voltage at $T$ = 1.9K (c) and at different temperatures at $V_g$

= -30 volts (d).

Continuing the analysis of our observations, figure 6a shows the change of conductance $\Delta\sigma_{xx}(B) = \sigma_{xx}(B)-\sigma_{xx}(B=0)$ of the ultra-thin WTe$_2$ device at $T = 1.9$K, after subtracting the classical contribution determined by the two-fluid model, and for the largest gate voltages ($V_g = \pm 80$ volts) applied in experiment. The low field magneto-conductance shows a peak that can be strongly modified by the gate voltage. Figure 6b shows the $\Delta\sigma_{xx}(B)$ curves for $T = 2$K, 12K and 30K, with $V_g = 30$ volts. It can be seen that the low field magneto-conductance diminishes rapidly as $T$ increases. Above 30K, the peak in low field magneto-conductance can no longer be detected. The magnitude and the temperature dependence of the low field magneto-conductance are characteristics of two-dimensional weak anti-localization, similar to previously reported result in ultra-thin WTe$_2$ devices [15, 19]. We fit our experimental data to the Hikami-Larkin-Nagaoka (HLN) equation [44]:

$$\Delta\sigma_{xx}(B) = -\frac{e^2}{2\pi^2\hbar}\left[\psi\left(\frac{1}{2}+\frac{B_\phi}{B}\right) - \ln\left(\frac{B_\phi}{B}\right) - 2\psi\left(\frac{1}{2}+\frac{B_\phi+B_{so}}{B}\right) + 2\ln\left(\frac{B_\phi+B_{so}}{B}\right) - \psi\left(\frac{1}{2}+\frac{B_\phi+2B_{so}}{B}\right) + \ln\left(\frac{B_\phi+2B_{so}}{B}\right)\right]$$

(4)

where $\psi$ is the digamma function, $B_n = \frac{\hbar}{4el_n^2} = \frac{\hbar}{4eD\tau_n}, n = \phi, so$ is the characteristic field related to phase coherence length (time) $l_\phi$ ($\tau_\phi$) and spin-orbit interaction terms. $D$ is the diffusion constant. In small magnetic field ($B<1$T), we find $B_{so}$ is too large to affect $\Delta\sigma_{xx}(B)$. Hence, following the literature, we fitted $\Delta\sigma_{xx}(B)$ curves by setting $B_{so}=6$T [19]. Figure 6c shows the dependence of the fitting parameter $B_\phi$ on $V_g$ at T=1.9K. It can be seen that $B_\phi$ drops by 50% if $V_g$ changes from 80 volts to -80 volts, indicating a significant increase in the phase coherence length of charge carriers as the sample tends to charge neutrality, consistent with the magnetoresistance data at higher magnetic field. $B_\phi$ increases linearly with temperature, which could be a manifestation of strong electron-electron interaction in the material [19, 45, 46].

**Conclusion**
In summary, we have fabricated ultra-thin WTe$_2$ field effect devices with solid gate dielectrics, and found that in electron-dominated regime, ultra-thin WTe$_2$ samples have a gate tunable magnetoresistance that is consistent with the two-fluid model. We estimate that the value of $\Delta MR/MR$ could be as high as 400,000% within experimentally accessible parameters, a value much higher than other materials. The tunability of $MR$ by a single parameter (the net charge density $n=n_e-n_h$) together with the insensitivity of $\Delta MR/MR$ to magnetic field and temperature, reveal the potential of ultra-thin WTe$_2$ as electric-field-tuned magnetoresistance material which could have important application in magnetic field sensing, information storage and extraction, and galvanic isolation.

**Acknowledgement**

This project has been supported by the National Basic Research Program of China (973 Grant Nos. 2013CB921900, 2014CB920900), and the National Natural Science Foundation of China (NSFC Grant Nos. 11374021) (X. Liu, Z. Zhang, C. Cai, S. Tian, H. Lu, S. Jia, J.-H. Chen). K.W. and T.T. acknowledge support from the Elemental Strategy Initiative conducted by the MEXT, Japan and a Grant-in-Aid for Scientific Research on Innovative Areas "Science of Atomic Layers" from JSPS. The work at Princeton University was supported by the ARO MURI on topological insulators, grant W911NF-12-1-0461.We are grateful to Professor Alberto Morpurgo for inspiring discussions.

**Author Contributions**

X.L. and J.C. conceived the experiment. X.L. exfoliated the $WTe_2$ thin flakes and few-layer h-BN crystals and accomplished the heterojunction FET devices fabrication. S.K, H.L., S.J. and R. J. C. grew $WTe_2$ bulk crystals; T.T. and K.W. grew h-BN bulk crystals. X.L., C.C and S.T. performed the transport measurements and Raman measurements. X.L., C.C, Z.Z. and J.C. discussed the results and analyzed the data. X.L. and J.C. wrote the manuscript and all authors commented on it.

# Supporting Information for

# Gate Tunable Magneto-resistance of Ultra-Thin WTe$_2$ Devices

Xin Liu[1*], Zhiran Zhang[1], Chaoyi Cai[1], Shibing Tian[1], Satya Kushwaha[2], Hong Lu[1], Takashi Taniguchi[3], Kenji Watanabe[3], Robert J. Cava[2], Shuang Jia[1,4*], Jian-Hao Chen[1,4*]

[1] International Center for Quantum Materials, Peking University, Beijing 100871, China
[2] Department of Chemistry, Princeton University, Princeton, USA
[3] High Pressure Group, National Institute for Materials Science, 1-1 Namiki, Tsukuba, Ibaraki 305-0044, Japan
[4] Collaborative Innovation Center of Quantum Matter, Beijing 100871, China

* Xin Liu (liux23@pku.edu.cn), Shuang Jia (jiashuang@gmail.com) and Jian-Hao Chen (chenjianhao@pku.edu.cn)

## S1. Characteristics of a Bulk WTe$_2$ Crystal.

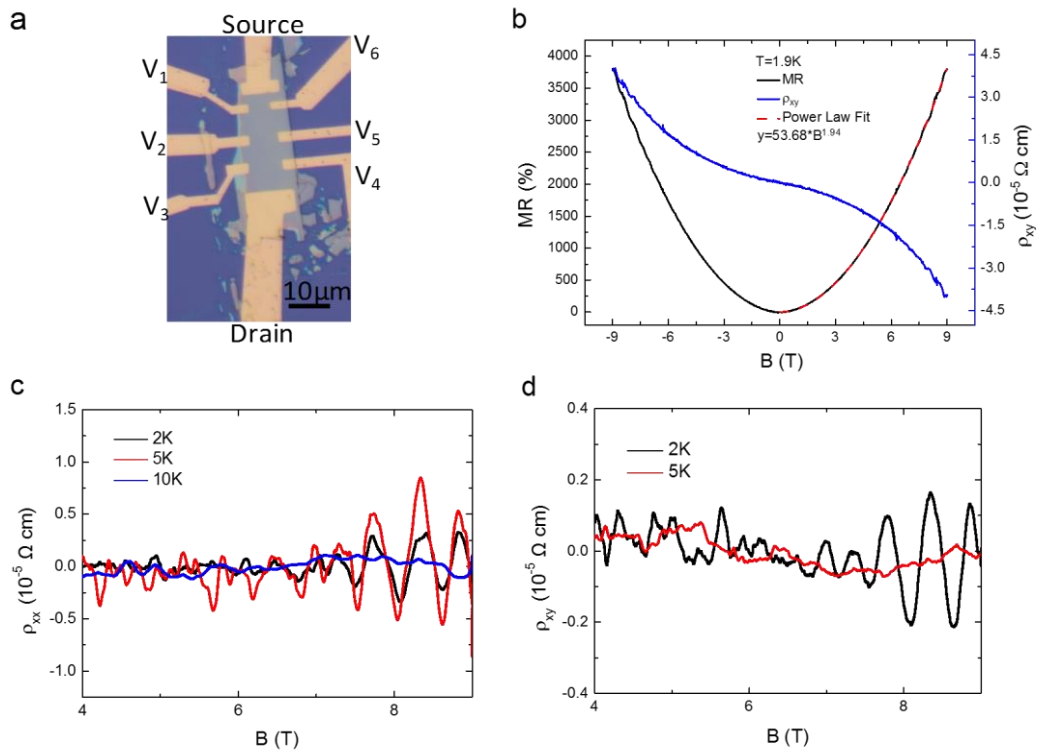

Figure S1a shows the optical image of ~112nm-thick WTe$_2$ FET device (sample B) which is mechanically exfoliated onto Si substrate with 300nm SiO$_2$. The source-drain current flows along the a-axis and the magnetic field is parallel to the c-axis. Longitudinal resistance $\rho_{xx}$ is measured with probes $V_1$ and $V_3$, while Hall resistance $\rho_{xy}$ is measured with probes $V_1$ and $V_6$. Magnetic field dependence and Hall resistance of the 112nm-thick WTe$_2$ crystal are shown in figure S1b. The MR of the measured device is up to 4000% under applied magnetic field of 9T and the residual resistivity ratio (RRR) is about 58.39 at B=0T. Furthermore, we make a power law fit to MR and the exponent is about 1.94[1, 2]. Shubnikov–de Haas oscillations are observed both in $\rho_{xx}$ and $\rho_{xy}$ shown in figure S1c and S1d, which suggests that the quality of the thicker sample is good. The colors mark different temperatures.

## S2. Temperature Dependence of Mobility in Bulk WTe$_2$ Crystals.

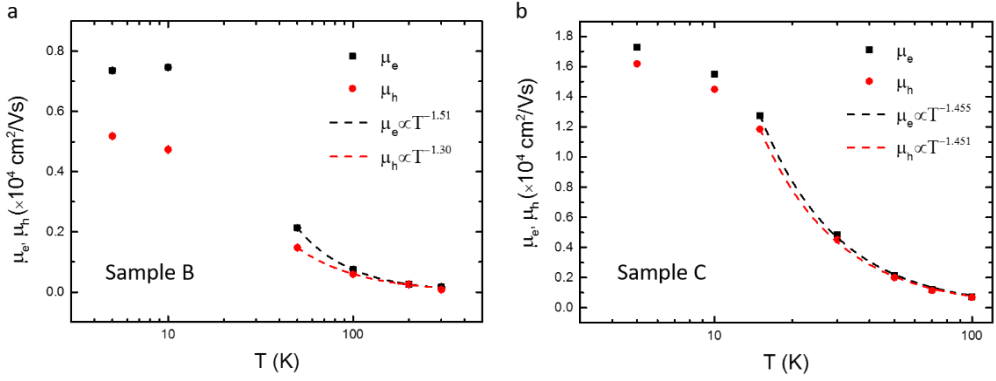

Figure S2 shows the dependence of mobilities $u_e$ and $u_h$ of sample B and C, respectively, as a function of temperature. Sample C is about 120nm thick and exfoliated from bulk WTe$_2$ crystal, whose growth process is different from sample B. Both mobilities for electron and hole of sample B in panel S2a increase following a power law fit of $\mu \propto T^{-\alpha}$ from 300K to 50K (100K to 15K, sample B in panel S2b) and then saturate from 10K to 2K (10K to 2K, sample B in panel S2a). A fit to a power law behavior of the decreasing mobility with increasing temperature give a larger exponent $\alpha$ for electrons ($\alpha$=1.51, 1.455) and holes ($\alpha$=1.30, 1.451) than WTe$_2$ thin flakes ($\alpha$=0.61 for electrons and $\alpha$=0.48 for holes) in the main text. Such values are close to the theoretically predicted value for bulk samples $\alpha$～1.52.

A lower $\alpha$ in ultra-thin WTe$_2$ crystals means that such samples could preserve their low temperature mobilities much better than their bulk counterparts at higher temperatures. The two-fluid model provides a strong correlation between higher charge carrier mobility and larger magnitude of the XMR of the samples. Such correlation is also observed in experiments on bulk WTe$_2$ [3]. Thus a lower $\alpha$ at ultra-thin WTe$_2$ means that the ultra-thin samples could preserve their response to magnetic field much better than their bulk counterparts at room temperature.

## S3. Raman Spectra of the Few-layer and Bulk WTe$_2$ Devices.

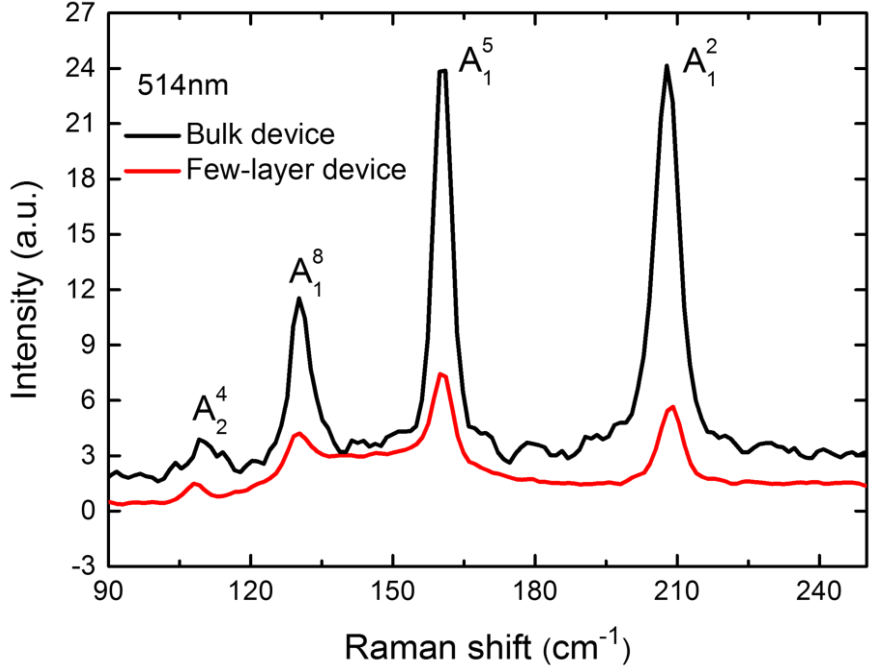

Figure S2 shows Raman spectra of the few-layer (sample A, in the main text) and 112nm-thick $WTe_2$ devices (sample B). Measurements are performed in a Horiba Jobin Yvon LabRam HR Evolution system. And we choose a non-polorized laser with wavelength of 514 nm at room temperature after accomplishing the transport experiments. Four prominent peaks are observed whose positions are $A_1^2$=209.14, $A_1^5$=159.96, $A_1^8$=130.306 and $A_2^4$= 107.995, respectively. The position in few-layer $WTe_2$ device is close to that in bulk materials (whose positions are $A_1^2$=207.8, $A_1^5$=159.845, $A_1^8$=130.19 and $A_2^4$= 109.12). The four Raman peaks suggest the measured $WTe_2$ samples are in 1T' phase [4] and it has not degraded.

**S4. Gate Dependence of Mobility in $WTe_2$ Thin Flakes on Different Substrate.**

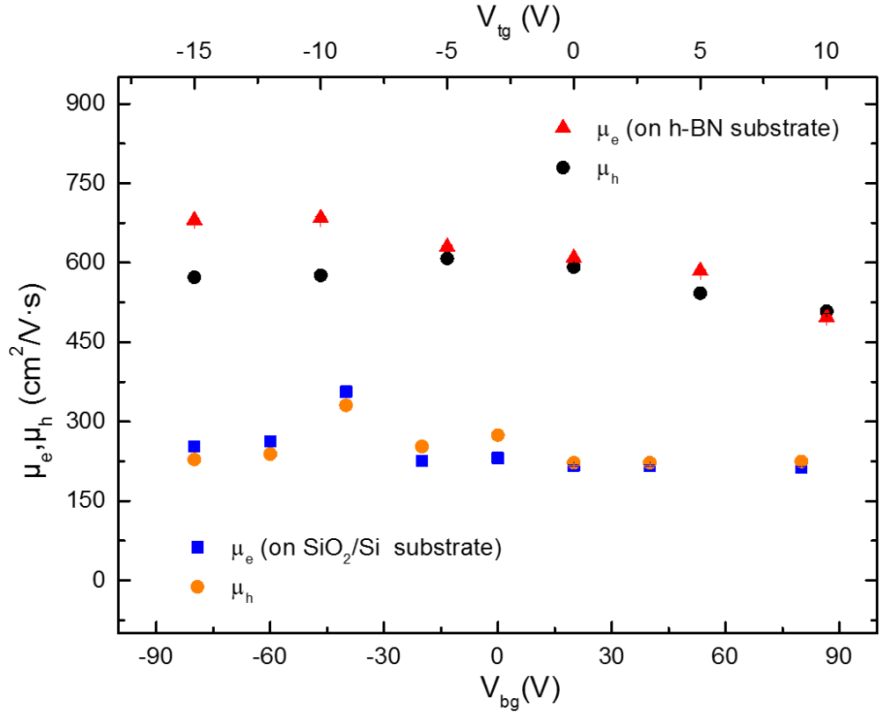

The gate dependence of mobility in WTe$_2$ devices (sample D and sample E) at T=2K is shown in Figure S3. WTe$_2$ thin flakes are mechanically exfoliated from bulk WTe$_2$ crystal and transferred on to thin h-BN single crystals (sample D) and SiO$_2$/Si substrates (sample E), respectively. Sample D is about 6-nm thick and sample E on the SiO$_2$/Si substrates is about 10-nm thick. There is an apparent increase in the mobility in device on h-BN substrate. In addition, better contact is achieved in thinner WTe$_2$ crystals on h-BN single crystal than on SiO$_2$/Si substrates.

**S5. Gate Dependence of $\rho_{xx}$ under Different In-plane Magnetic Fields.**

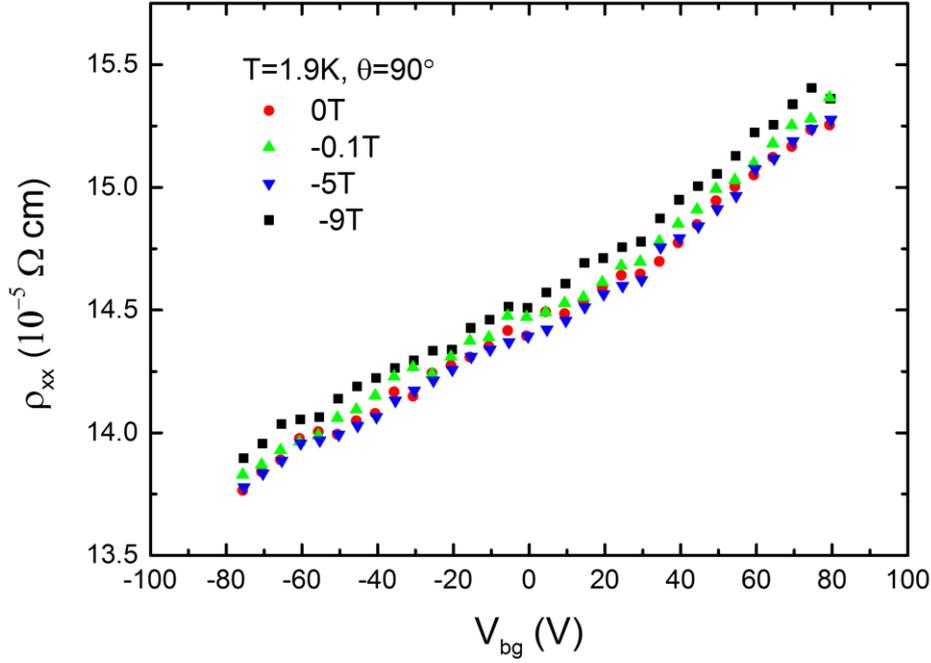

The gate dependence of longitudinal resistance of sample A at temperature of 1.9 Kelvin under various in-plane magnetic fields is shown in Figure S3. We find very little MR in the in-plane magnetic field, which is similar to that in bulk materials [5].

**S6. Temperature Dependence of Longitudinal Resistance.**

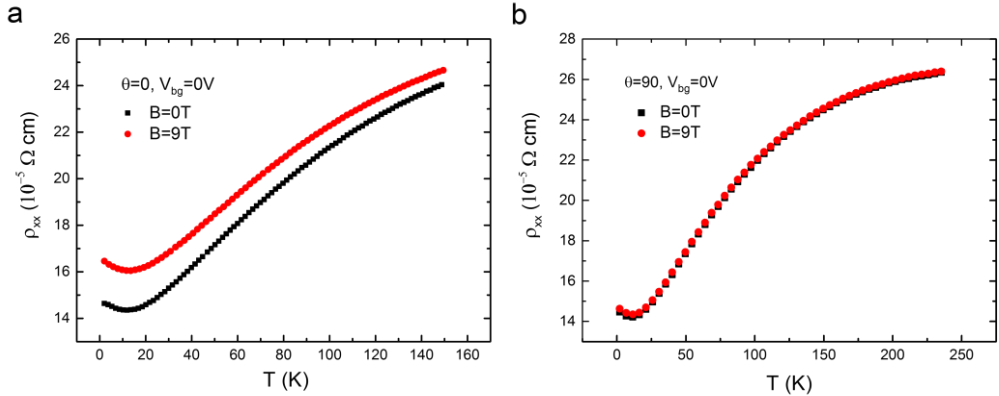

Figure S4 shows temperature dependence of longitudinal resistance of sample A under 0T and 9T magnetic field with zero back gate voltage. In figure S4a, magnetic field is applied perpendicular to the a-axis. We see a small increase in resistivity below 12K at zero magnetic field for our device, and the application of 9T magnetic field does not move the turn-on temperature T* much higher, which is likely caused by the suppression of MR in few-layer $WTe_2$ crystals. When the magnetic field is parallel to the ab plane (Fig. S4b), the MR is minimal for the whole temperature range of 1.9K to 250K.

**S7. Magnetic Field Dependence of Power-Law-Fit Exponent**

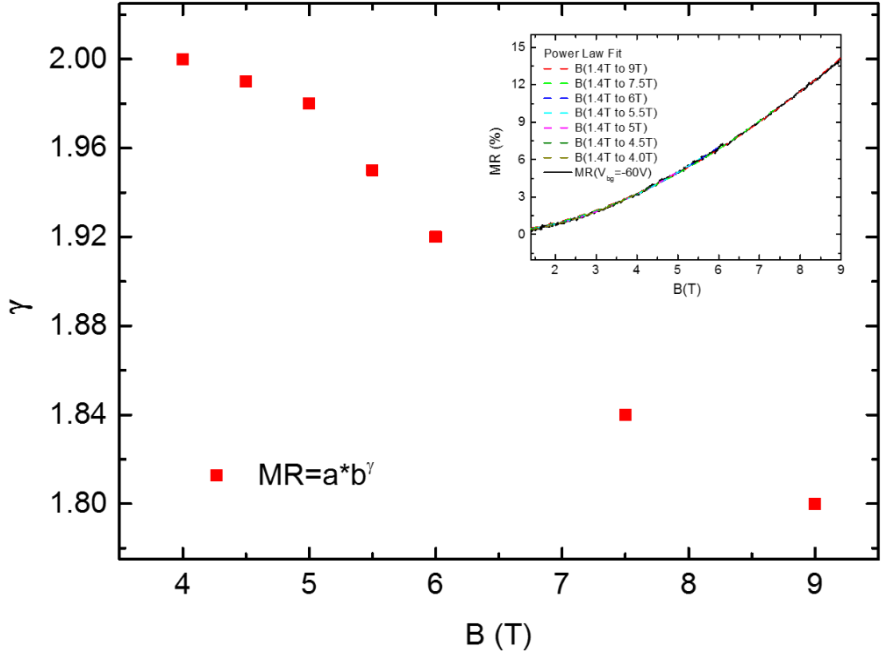

Figure S7 shows the power law fitting $MR \sim B^\gamma$ at different range of magnetic field for an experimental MR curve at back gate voltage -60V. It is seen in the inset panel that by fitting the MR between 1.4T and 4.0T, we obtained an exponent gamma=2; for MR between 1.4T and 6T, we obtained gamma=1.92; for MR between 1.4T and 9T, we obtained gamma=1.8. The fitting shows that the exponent $\gamma$ is not a constant of magnetic field, and reveals the trend of saturation of MR at high enough magnetic field. Furthermore, this shows that the simple mathematical expression ($MR \sim B^\gamma$) might not capture the physics and we use the two fluid model to fit and understand the experimental data in the main text.

## S8. Linear Extrapolation of Carrier Density and Mobility

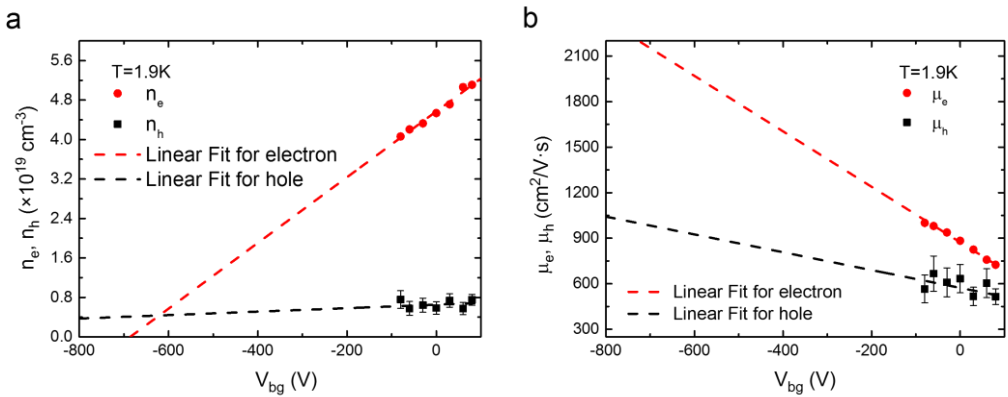

Figure S8 shows carrier density and mobility as a function of back gate voltage at temperature of 1.9 Kelvin. Dashed lines are linear fit to the experimental data, with specific parameters shown in equation (S1) to (S4). The linearity of our data is very good in the experimentally explored regime. We extend the linear extrapolation in $n_e$ and $n_h$ vs $V_g$ and find that neutrality will be obtained at $V_{bg}$=-620V (Figure S8a). To estimate the MR at neutrality for our device, we extend the linear extrapolation in $\mu_e$ and $\mu_h$ vs $V_g$ at $V_{bg}$=-620V and obtain a change of MR to be 8,400%.

$$n_e = 6.67 \times 10^{-3} + 4.57 V_{bg} \tag{1}$$
$$n_h = 3.52 \times 10^{-4} + 0.65 V_{bg} \tag{2}$$
$$\mu_e = 873.26 - 1.83 V_{bg} \tag{3}$$
$$\mu_h = 572.43 - 0.58 V_{bg} \tag{4}$$

If the mobility of pristine bulk sample (as high as 167,000 cm$^2$V$^{-1}$s$^{-1}$ from ref.[3]) can be preserved, when the carrier density is adjusted from p/n=0.1599 to p/n=1, we use the constant mobility value (167,000 cm$^2$V$^{-1}$s$^{-1}$) to obtain a lower bound for the change of MR in such device to be 402,600%. Such rapid improvement of device fabrication techniques has been seen in the field of graphene, where it did not take a long time for the mobilities of graphene devices to improve from 10,000cm$^2$/Vs [6] to 1,000,000cm$^2$/Vs [7].

### S9. Simultaneous Fitting of $\rho_{xx}$ and $\rho_{xy}$ Curves with the Least Squares Method

The two-fluid model gives:

$$\rho_{xx} = \frac{1}{e} \frac{(n_e u_e + n_h u_h) + (n_e u_h + n_h u_e) u_e u_h B^2}{(n_e u_e + n_h u_h)^2 + ((n_e - n_h) u_e u_h B)^2} \tag{5}$$

$$\rho_{xy} = \frac{1}{e} \frac{(n_e \mu_e^2 - n_h \mu_h^2) - (n_h - n_e) \mu_e^2 \mu_h^2 B^2}{(n_e u_e + n_h u_h)^2 + ((n_h - n_e) u_e u_h B)^2} B \tag{6}$$

where $n_e$ ($n_h$) and $u_e$ ($u_h$) are carrier density and mobility for electrons (holes), respectively. Here, we need to minimize the goodness-of-fit of between the experimental data of $\rho_{xx}$ ($\rho_{xy}$) and theoretical curve from equation (S5) ((S6)) at each gate voltage and temperature.

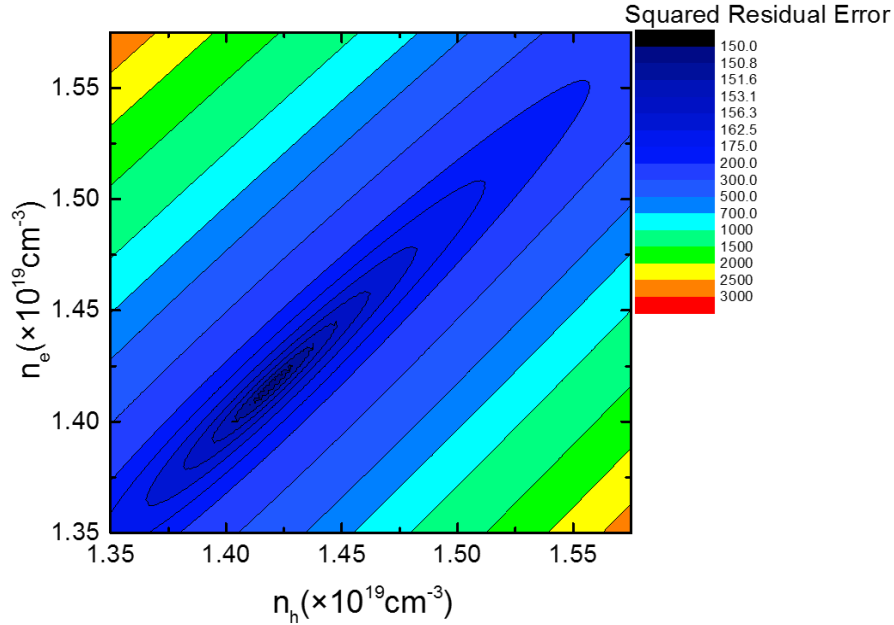

We first calculate the squared error of $\rho_{xx}$ at a specific gate voltage and temperature. Since it is a function of four parameters, $n_e$, $n_h$, $u_e$, $u_h$, which could not be plotted in a straightforward way, we first determine the least squared error surface in this 4-dimensional parameter space with the optimal $\mu_h$ and $\mu_e$ for each ($n_h$, $n_e$); e.g., we minimize the error for each point of ($n_h$, $n_e$) with the best $\mu_h$ and $\mu_e$. Such least squared error surface could be drawn with $n_h$ (X axis) and $n_e$(Y axis) and the squared error (Z axis). Then we use the same set of $n_e$, $n_h$, $u_e$, $u_h$ to obtain the least squared error surface of $\rho_{xy}$. The joint least squared error of $\rho_{xx}$ and $\rho_{xy}$ versus $n_h$ and $n_e$ will be graphs similar to figure S6 (Figure S9 is obtained for $V_g = 0V$ and $T = 2K$). Now we can determined the optimal $n_h$, $n_e$, $\mu_h$ and $\mu_e$ for each $V_g$ and $T$ from the point of such graphs that has the minimum squared error.